\documentclass[pra,
                preprint,
                amsmath,
                amssymb,
                showpacs,
                superscriptaddress]{revtex4}
\usepackage{graphicx} % Include figure files
\usepackage{dcolumn}  % Align table columns on decimal point
\usepackage{bm}       % bold math
\usepackage{amsfonts}
\usepackage{dsfont}
\begin{document}

\title{Theory of time-resolved non-resonant x-ray scattering for imaging ultrafast coherent electron motion}

\author{Gopal Dixit}
\email[]{dixit@mbi-berlin.de}
\affiliation{%
Center for Free-Electron Laser Science, DESY,
            Notkestrasse 85, 22607 Hamburg, Germany }
\affiliation{%
The Hamburg Centre for Ultrafast Imaging,
Luruper Chaussee 149, 22761 Hamburg, Germany }

\author{Jan Malte Slowik}
\affiliation{%
Center for Free-Electron Laser Science, DESY,
            Notkestrasse 85, 22607 Hamburg, Germany }
\affiliation{%
The Hamburg Centre for Ultrafast Imaging,
Luruper Chaussee 149, 22761 Hamburg, Germany }
\affiliation{%
Department of Physics, University of Hamburg, 20355 Hamburg,
Germany}

\author{Robin Santra}
\email[]{robin.santra@cfel.de}
\affiliation{%
Center for Free-Electron Laser Science, DESY,
            Notkestrasse 85, 22607 Hamburg, Germany }
\affiliation{%
The Hamburg Centre for Ultrafast Imaging,
Luruper Chaussee 149, 22761 Hamburg, Germany }
\affiliation{%
Department of Physics, University of Hamburg, 20355 Hamburg,
Germany}

\date{\today}

\pacs{34.50.-s, 61.05.cf, 78.70.Ck}

%%%%%%%%%%%%%%%%% END OF PREAMBLE %%%%%%%%%%%%%%%%

\begin{abstract}
Future ultrafast x-ray light sources might image ultrafast coherent electron motion in
real-space and in real-time. 
For a rigorous understanding of such an imaging experiment, we extend the theory of non-resonant x-ray scattering to the time-domain.
The role of energy resolution of the scattering detector is investigated in detail.
We show that time-resolved non-resonant x-ray scattering with no energy resolution offers an opportunity to study
time-dependent  electronic correlations in non-equilibrium quantum systems. 
Furthermore, our theory presents a unified description of ultrafast x-ray scattering from electronic wave packets and the dynamical 
imaging of ultrafast dynamics using inelastic x-ray scattering by Abbamonte and co-workers. 
We examine closely the relation of the scattering signal and the linear density response of electronic wave packets.
Finally, we demonstrate that time-resolved x-ray scattering from a crystal consisting of identical electronic wave packets recovers the instantaneous electron density.
\end{abstract}

\maketitle
\section{Introduction}
Scattering of x rays from matter is a well-established method in
several areas of science to access real-space, atomic-scale
structural information of complex materials, ranging from
molecules to biological complexes~\cite{Ihee, Chapman,
Seibert}. Utilizing the Fourier relationship between the electron  
density of the sample and the scattering intensity (i.e., elastic x-ray scattering),
coherent diffractive imaging (CDI) is a powerful lensless technique to obtain 
three dimensional structural information of
non-periodic and periodic samples~\cite{chapman2010, abbey2011, miao1999, zuo2003}.
With the recent progress in technology for producing ultrashort, tunable,
and high-energy x-ray pulses from x-ray free-electron lasers
(XFELs)~\cite{emma2, ishikawa2012}, a particular interest has been
aroused to perform CDI with atomic-scale spatial resolution at
present and forthcoming XFELs (LCLS, SACLA, European XFEL). In
addition, the high brightness of the x-ray pulses from XFELs
promises the possibility to carry out single-shot CDI with
sufficiently strong scattering signal for imaging individual non-periodic
objects. 

The natural timescale of electronic motion ranges from tens of
attoseconds (1 as = 10$^{-18}$ s) to few femtoseconds (1 fs =
10$^{-15}$ s)~\cite{krausz, bucksbaum2007, corkum2007}. In order
to understand how spatial properties of electronic states change
in time, it is crucial to image the dynamical evolution of the
electronic charge distribution with angstrom spatial
resolution and (sub-)fs temporal resolution. Hence, imaging the
electronic charge distribution with atomic-scale spatial and
temporal resolutions will provide a unique opportunity to
understand several ubiquitous ultrafast phenomena like
electron-hole dynamics and electron transfer
processes~\cite{breidbach2003, kuleff2005, remacle, dutoi2011}.
The pump-probe approach is one of the most common ways to study
ultrafast dynamics, where first a pump pulse
activates the dynamics and subsequently the activated dynamics are
investigated by the probe pulse at a precise instant. Recently,
the synchronization to within tens of attoseconds between pump and
probe pulses has been demonstrated
experimentally~\cite{benedick2012}. Moreover, attosecond hard
x-ray pulses seem feasible in the near future~\cite{Emma1,
Zholents, tanaka2013, kumar2013attosecond}. Therefore, the x-ray
pulses will be comparable to the natural timescale of several
elementary processes in nature and will open the door to study
these ultrafast processes in real-space and in real-time.

Time-resolved \mbox{x-ray} scattering (TRXS) from temporally
evolving electronic systems is an emerging and promising approach
for real-time and real-space imaging of the electronic motion. A
series of scattering patterns obtained at different instants of
the dynamics may be stitched together to make a movie of the
electronic motion with unprecedented spatiotemporal resolution. In
this context, a straightforward extension of x-ray scattering from the static to the time
domain would seem to suggest the possibility of imaging ultrafast electronic motion with the notion
that the scattering pattern encodes information related to  the
instantaneous electron density.

In order to image the electronic motion on an ultrafast timescale, the probe pulse duration must be smaller than the
characteristic timescale of the motion. As a consequence, the ultrashort probe
pulse has a finite, broad spectral bandwidth. 
Thus, it is fundamentally difficult
to perform an energy-resolved scattering experiment with an energy
resolution better than the bandwidth of the pulse. 
This makes it
necessary to include all transitions within the bandwidth induced
by the scattering process. 
In our previous work, we have focused on the imaging of coherent electronic motion in a hydrogen atom via
quasi-elastic TRXS assuming high energy resolution of the scattering detector~\cite{dixit2012}. 
Furthermore, we have also investigated the role of scattering interference 
between a non-stationary and several stationary electrons in a  many-electron system. 
The findings of the scattering interference were visually demonstrated for the helium atom, 
where one electron forms an electronic wave packet and the other electron remains stationary~\cite{dixit2013jcp}.  
We also proposed time-resolved phase-contrast imaging as a future experiment to image instantaneous 
electron density~\cite{dixit2013prl}.

The purpose of the present paper is to provide a rigorous theoretical  analysis
of the imaging of coherent electronic motion via  TRXS and to discuss the pros and cons of TRXS. 
In this work, we provide a unified description of ultrafast x-ray scattering from electronic wave packets and the dynamical 
imaging of  ultrafast dynamics using inelastic x-ray scattering introduced by 
Abbamonte and co-workers~\cite{abbamonte2004imaging, abbamonte2008dynamical, abbamonte2010ultrafast, abbamonte2009inhomogeneous, abbamonte2012standingwaves}. 
This paper is structured as follows. Section II discusses the theory of TRXS from electronic wave packets. 
Section III presents results and a discussion of the theory presented in the previous section. 
Section III is sub-divided into three subsections, where we present: 
A) the role of the energy resolution of the scattering detector, especially the cases of no and high energy resolution;
B) the density perturbation response of electronic wave packets within linear response theory;   
and 
C) TRXS from  a crystal consisting of identical electronic wave packets at each lattice point. 
Conclusions are presented  in Sec. IV. 
The detailed mathematical steps are presented in the three appendices.   

\section{Theory}
\label{secII}

Atomic units are used throughout this article unless specified otherwise. 
We begin with the minimal-coupling interaction
Hamiltonian for light-matter interaction in the Coulomb gauge
~\cite{craig1984}
\begin{equation}\label{eq1}
\hat{H}_{\textrm{int}} = \alpha \int d^{3}x
~\hat{\psi}^{\dagger}(\mathbf{x})~
\left[\hat{\mathbf{A}}(\mathbf{x}) \cdot
\frac{\boldsymbol{\nabla}}{i}\right] ~\hat{\psi}(\mathbf{x}) +
\frac{\alpha^{2}}{2}\int d^{3}x ~\hat{\psi}^{\dagger}(\mathbf{x})~
\hat{\mathbf{A}}^{2} (\mathbf{x})~\hat{\psi}(\mathbf{x}),
\end{equation}
where $\alpha$ is the fine-structure constant,
$\hat{\psi}^{\dagger}(\mathbf{x}) ~ [\hat{\psi}(\mathbf{x})]$ is the
creation (annihilation) field operator for an electron at position
$\mathbf{x}$, $\hat{\mathbf{A}}$ is the vector potential operator of the
light and $\frac{\boldsymbol{\nabla}}{i}$ is the canonical
momentum of an electron. It is well established that at photon
energies much higher than all inner-shell thresholds in the system
of interest, elastic and inelastic scattering (Thomson and Compton
scattering) are mediated by the $\hat{\mathbf{A}}^{2} $ operator.
Therefore, we only focus on scattering mediated by
$\hat{\mathbf{A}}^{2} $ and will not consider the contribution
from the dispersion correction in the scattering process, i.e.,
scattering induced by the  $\hat{\mathbf{A}}(\mathbf{x}) \cdot \boldsymbol{\nabla}$ operator  in second order.
In the inelastic case, the $\hat{\mathbf{A}}^{2} $ induced
scattering is also known as non-resonant inelastic x-ray
scattering~\cite{haverkort2007nonresonant}. Most generally, the x rays must be
treated as a statistical mixture of photons occupying all possible
electromagnetic modes. $\hat{\mathbf{A}}$ can be expressed in
terms of plane waves as~\cite{craig1984}
\begin{equation}\label{eq2}
\hat{\mathbf{A}}(\mathbf{x}) = \sum_{\mathbf{k}, s} \sqrt{\frac{2
\pi}{V \omega_{\mathbf{k}} \alpha^{2}}} \left\{
{\hat{a}_{\mathbf{k}, s}} \boldsymbol{\epsilon}_{\mathbf{k}, s}
e^{i \mathbf{k} \cdot \mathbf{x}} + \hat{a}^{\dagger}_{\mathbf{k},
s} \boldsymbol{\epsilon}^{*}_{\mathbf{k}, s} e^{-i \mathbf{k}
\cdot \mathbf{x}} \right \},
\end{equation}
where $V$ is the quantization volume, $\omega_{\mathbf{k}}$ is the
energy of a photon in the $\mathbf{k}$-th mode, $\mathbf{k}$ and
$s$ are the wave vector and the polarization index of a given
mode, respectively. $\hat{a}^{\dagger}_{\mathbf{k},s} ~
(\hat{a}_{\mathbf{k},s})$ is the photon creation (annihilation)
operator and $\boldsymbol{\epsilon}_{\mathbf{k}, s}$ is the
polarization vector in the $\mathbf{k}, s$ mode.

Here, we assume that an electronic wave packet  $| \Phi, t \rangle$ 
has been prepared with the help of a suitable pump pulse
with sufficiently broad energy bandwidth. 
To obtain the differential scattering probability (DSP), which is the crucial quantity in
x-ray scattering,
we employ first-order time-dependent perturbation theory to the
interaction between matter and x rays. The expression for the DSP
is~\cite{dixit2012}
\begin{eqnarray}\label{eq3}
\frac{dP}{d\Omega} & = &
\frac{d\sigma_{th}}{d\Omega}~\int_{-\infty}^{\infty} d\tau
\int_{-\infty}^{\infty} d\delta \int_{0}^{\infty}
d\omega_{\mathbf{k}_{s}} ~W_{\Delta E}
({\omega_{\mathbf{k}_{s}}}) \frac{\omega_{\mathbf{k}_{s}}
}{(2\pi \omega_{\mathbf{k}_{in}})^{2}\alpha}
~ e^{ - i\omega_{\mathbf{k}_{s}} \delta } \nonumber \\
& & \times \int d^{3}x ~ \int d^{3}x^{\prime} \left \langle \Phi
\left| ~\hat{n} \left( \mathbf{x}^{\prime}, \tau +
\frac{\delta}{2} \right)~\hat{n} \left(\mathbf{x}, \tau -
\frac{\delta}{2} \right)
~ \right| \Phi \right \rangle  \nonumber \\
&& \times e^{- i \mathbf{k}_{s} \cdot (
\mathbf{x}-\mathbf{x}^{\prime})}~ G^{(1)}
\left(\mathbf{x}^{\prime}, \tau+\frac{\delta}{2};~ \mathbf{x},
\tau-\frac{\delta}{2} \right),
\end{eqnarray}
where $\frac{d\sigma_{th}}{d\Omega}$ is the Thomson scattering
cross section, $\omega_{\mathbf{k}_{in}}$ is the photon energy of the incident central carrier frequency
and $\omega_{\mathbf{k}_{s}}$ refers to the scattered photon energy. $\mathbf{k}_{s}$ is the momentum
of the scattered photon, $\hat{n}(\mathbf{x})$ is the electron
density operator, and $G^{(1)}$ is the first-order correlation function for
the x rays~\cite{loudon1983, glauber1963}. The energy resolution of the
scattering detector is specified by a spectral window function,
$W_{\Delta E}({\omega_{\mathbf{k}_{s}}})$, which is a function of
$\omega_{\mathbf{k}_{s}}$ with a width $\Delta E$ modeling the
range of scattered photon energies accepted by the detector. It is
important to note that the window function is not a normalized
function, i.e., the detected scattering intensity depends on the width of
the window function, which implies that the signal is weak for
narrow width $\Delta E$.

We assume, for simplicity, that the x rays can be treated as a coherent ensemble of Gaussian pulses; 
the expression for the first-order correlation function is given in
Appendix \ref{app:1}. 
Furthermore, we assume the object much smaller than the distance $c\tau_{l}$, 
where $c$ is the speed of light and $\tau_{l}$ the pulse duration.
Now, the DSP from Eq.~(\ref{eq3}) reduces  to
\begin{eqnarray}\label{eq4}
\frac{dP}{d\Omega} & = &  \frac{d\sigma_{th}}{d\Omega} \int_{0}^{\infty}
d\omega_{\mathbf{k}_{s}} ~ W_{\Delta E}
({\omega_{\mathbf{k}_{s}}}) ~
\frac{\omega_{\mathbf{k}_{s}}}{\omega_{\mathbf{k}_{in}}} ~
\int_{-\infty}^{\infty} d\tau \frac{I(\tau)} {\omega_{\mathbf{k}_{in}}}
\int_{-\infty}^{\infty} \frac{d\delta}{2\pi} ~C(\delta) ~ e^{-
i(\omega_{\mathbf{k}_{s}}-
\omega_{\mathbf{k}_{in}}) \delta }\nonumber \\
& & \times \int d^{3}x \int d^{3}x^{\prime} ~ \left \langle \Phi
\left| ~\hat{n}\left(\mathbf{x}^{\prime}, \tau + \frac{\delta}{2}
\right)~\hat{n}\left(\mathbf{x}, \tau - \frac{\delta}{2}\right)
\right| \Phi \right \rangle e^{i \mathbf{Q}\cdot(\mathbf{x} -
\mathbf{x}^{\prime})}.
\end{eqnarray}
Here, $I(\tau)$ is the intensity of the probe pulse, $C(\delta) =
\exp[{-2\ln{2}\;\delta^{2}/\tau_{l}^{2}}]$ is a function of the pulse duration
$\tau_{l}$, and $\mathbf{Q} = \mathbf{k}_{in} - \mathbf{k}_{s} $ is
the photon momentum transfer with $\mathbf{k}_{in}$ as the
incident photon momentum. Equation~(\ref{eq4}) is the key expression for
TRXS, but a straightforward interpretation of this equation is
not easy as it is a complicated expression of $\tau$, $\delta$ and
$\omega_{\mathbf{k}_{s}}$ variables. The electronic correlation function in  
Eq.~(\ref{eq4}) reflects the perturbation of the freely evolving
electronic wave packet by the density operator. 
Through this density perturbation electronic states can be populated that 
initially were not present in the wave packet. Furthermore, these
freely evolving, additionally populated electronic states get
projected back onto the density-perturbed wave packet at a later time [see second line in Eq.~(\ref{eq4})]. 
It is thus evident from Eq.~(\ref{eq4}) that TRXS is related
to an intricate quantity: a space-time dependent
density-density correlation function, which is in contrast with
the common notion that TRXS  provides access to the
instantaneous electron density $\langle \hat{n}(\mathbf{x})
\rangle_{t} = \rho(\mathbf{x}, t)$. A similar approach for ultrafast x-ray scattering has
been developed in the past~\cite{henriksen, tanaka}, but focused on x-ray scattering for probing
atomic motion, e.g., bond breaking in diatomic
molecules~\cite{henriksen}.

\section{Results and Discussion}
In the following, we will further elucidate Eq.~(\ref{eq4})
taking into consideration a key assumption for TRXS. The probe pulse is
assumed to be sufficiently short to freeze the wave packet
dynamics, i.e., the evolution of the wave packet is assumed to be
much slower than the pulse duration. Under this situation, the
$\tau$-dependent phases of the wave packet can be collected
together with the $I(\tau)$, and the $\tau$-dependent integral can
be performed in Eq.~(\ref{eq4}), yielding
\begin{eqnarray}\label{eq41}
\frac{dP}{d\Omega} & = &
\frac{d\sigma_{th}}{d\Omega}~\mathcal{F}~ \int_{0}^{\infty}
d\omega_{\mathbf{k}_{s}} ~ W_{\Delta E}
({\omega_{\mathbf{k}_{s}}}) ~
\frac{\omega_{\mathbf{k}_{s}}}{\omega_{\mathbf{k}_{in}}} ~
\int_{-\infty}^{\infty} \frac{d\delta}{2\pi} ~C(\delta) ~ e^{-
i(\omega_{\mathbf{k}_{s}}-
\omega_{\mathbf{k}_{in}}) \delta }\nonumber \\
& & \times \int d^{3}x \int d^{3}x^{\prime} ~ \left \langle \Phi
\Biggl| \hat{n}\left(\mathbf{x}^{\prime}, \tau_{d} +
\frac{\delta}{2} \right)~\hat{n}\left(\mathbf{x}, \tau_{d} -
\frac{\delta}{2}\right) \Biggr| \Phi \right \rangle e^{i
\mathbf{Q}\cdot(\mathbf{x} - \mathbf{x}^{\prime})}.
\end{eqnarray}
Here, $\mathcal{F}$ is the fluence of the probe pulse (in units of number of photons per area) and
$\tau_{d}$ is the pump-probe delay time. The above equation can be
re-written as
\begin{eqnarray}\label{eq42}
\frac{dP}{d\Omega} & = &
\frac{d\sigma_{th}}{d\Omega}~\mathcal{F}~\int_{0}^{\infty}
d\omega_{\mathbf{k}_{s}} ~ W_{\Delta E}
({\omega_{\mathbf{k}_{s}}}) ~
\frac{\omega_{\mathbf{k}_{s}}}{\omega_{\mathbf{k}_{in}}} ~
\int_{-\infty}^{\infty} \frac{d\delta}{2\pi} ~C(\delta) ~ e^{-
i(\omega_{\mathbf{k}_{s}}-
\omega_{\mathbf{k}_{in}}) \delta }\nonumber \\
& & \times \int d^{3}x \int d^{3}x^{\prime} ~ \left \langle \Phi
\Biggl| e^{i\hat{H} \frac{\delta}{2}}
\hat{n}\left(\mathbf{x}^{\prime}, \tau_{d} \right)
e^{-i\hat{H}\delta}~\hat{n}\left(\mathbf{x}, \tau_{d}\right)
e^{i\hat{H} \frac{\delta}{2}} \Biggr| \Phi \right \rangle e^{i
\mathbf{Q}\cdot(\mathbf{x} - \mathbf{x}^{\prime})}.
\end{eqnarray}
Here, $\hat{H}$ is the electronic Hamiltonian. Furthermore, by introducing  $\langle \hat{H} \rangle =
\tilde{E}$ as the mean energy of the wave packet, the
$\delta$-dependent freely evolving phase of the wave packet,
$\textrm{exp}[{i E_{i} {\delta}/{2}}]$, can be factorized into
$\textrm{exp}[{i  \tilde{E}{\delta}/{2}}]$ and $\textrm{exp}[{i
(E_{i}- \tilde{E} ){\delta}/{2}}]$ with $E_{i}$ as the eigen-energy corresponding to eigenstate $| \Psi_{i} \rangle$ in the
wave packet. Since the pulse duration is short enough to freeze
the motion, $|E_{i}- \tilde{E}| \ll 1/\delta$ holds. Therefore,
$\textrm{exp}[{i  (E_{i}- \tilde{E} ){\delta}/{2}}]$ can be
approximated by unity and Eq.~(\ref{eq42}) can be written as
\begin{eqnarray}\label{eq43}
\frac{dP}{d\Omega} & = &
\frac{d\sigma_{th}}{d\Omega}~\mathcal{F}~\int_{0}^{\infty}
d\omega_{\mathbf{k}_{s}} ~ W_{\Delta E}
({\omega_{\mathbf{k}_{s}}}) ~
\frac{\omega_{\mathbf{k}_{s}}}{\omega_{\mathbf{k}_{in}}} ~
\int_{-\infty}^{\infty} \frac{d\delta}{2\pi} ~C(\delta) ~ e^{-
i(\omega_{\mathbf{k}_{s}}-
\omega_{\mathbf{k}_{in}}) \delta }\nonumber \\
& & \times \int d^{3}x \int d^{3}x^{\prime} ~ \left \langle
\Phi\Biggl| \hat{n}\left(\mathbf{x}^{\prime}, \tau_{d} \right)
e^{-i ( \hat{H}-\tilde{E})\delta}~\hat{n}\left(\mathbf{x}, \tau_{d}
\right) \Biggr| \Phi \right \rangle e^{i
\mathbf{Q}\cdot(\mathbf{x} - \mathbf{x}^{\prime})}.
\end{eqnarray}
Now the $\delta$-dependent  integral can be performed straight
away in the above equation, which yields the simplified expression
for the DSP as
\begin{eqnarray}\label{eq44}
\frac{dP}{d\Omega} & = &
\frac{d\sigma_{th}}{d\Omega}~\mathcal{F}~ \int_{0}^{\infty}
d\omega_{\mathbf{k}_{s}} ~ W_{\Delta E}
({\omega_{\mathbf{k}_{s}}}) ~
\frac{\omega_{\mathbf{k}_{s}}}{\omega_{\mathbf{k}_{in}}}
\frac{\tau_{l}}{\sqrt{8 \pi \textrm{ln}2}} \nonumber \\
& & \times \int d^{3}x \int d^{3}x^{\prime} ~ \left \langle \Phi
\Biggl| \hat{n}\left(\mathbf{x}^{\prime}, \tau_{d} \right) e^{
-\frac{\tau^{2}_{l}}{8 \textrm{ln}2} ( \omega_{\mathbf{k}_{in}}-
{\omega_{\mathbf{k}_{s}}} +\tilde{E}
-\hat{H})^{2}} ~\hat{n}\left(\mathbf{x}, \tau_{d} \right) \Biggr|
\Phi \right \rangle e^{i \mathbf{Q}\cdot(\mathbf{x} -
\mathbf{x}^{\prime})}.
\end{eqnarray}
Let us introduce a complete set of eigenstates in between the
two density operators in Eq.~(\ref{eq44}), such that 
\begin{eqnarray}\label{eq441}
\frac{dP}{d\Omega} & = & \frac{d\sigma_{th}}{d\Omega}~\mathcal{F}~
\int_{0}^{\infty} d\omega_{\mathbf{k}_{s}} ~ W_{\Delta E}
({\omega_{\mathbf{k}_{s}}}) ~
\frac{\omega_{\mathbf{k}_{s}}}{\omega_{\mathbf{k}_{in}}}
\sum_{f} \frac{\tau_{l}}{\sqrt{8 \pi \textrm{ln}2}} e^{
-\frac{\tau^{2}_{l}}{8 \textrm{ln}2} ( \omega_{\mathbf{k}_{in}} - {\omega_{\mathbf{k}_{s}}}
+ \tilde{E} - E_{f})^{2}} \nonumber \\
& & \times \int d^{3}x \int d^{3}x^{\prime} ~ \langle
\Phi|\hat{n}(\mathbf{x}^{\prime}, \tau_{d}) |\Psi_{f} \rangle ~
\langle \Psi_{f} | \hat{n}(\mathbf{x}, \tau_{d}) | \Phi \rangle
e^{i \mathbf{Q}\cdot(\mathbf{x} - \mathbf{x}^{\prime})}.
\end{eqnarray}
Here, $|\Psi_{f}\rangle$ and $E_{f}$ are the electronic state reached by x-ray scattering
and the associated electronic energy, respectively. 

At this point, it is instructive to recover from TRXS the case of x-ray scattering from a stationary target.
To image a stationary electronic state the pulse duration may become arbitrarily large. 
Considering the monochromatic limit $\tau_{l} \rightarrow \infty$, one obtains from Eq.~(\ref{eq4}) the general expression for x-ray scattering from a stationary target  
\begin{eqnarray}\label{eq5}
\frac{dP}{d\Omega} & = & \frac{d\sigma_{th}}{d\Omega}~\mathcal{F}~
\int_{0}^{\infty} d\omega_{\mathbf{k}_{s}}~W_{\Delta E}
({\omega_{\mathbf{k}_{s}}}) ~
\frac{\omega_{\mathbf{k}_{s}}}{\omega_{\mathbf{k}_{in}}}
\sum_{f} \delta( \omega_{\mathbf{k}_{in}} - \omega_{\mathbf{k}_{s}} + E_{0} - E_{f}) \nonumber \\
& & \times \int d^{3}x \int d^{3}x^{\prime} ~
\langle \Psi_{0} | \hat{n} (\mathbf{x}^{\prime}) | \Psi_{f} \rangle ~
\langle \Psi_{f} | \hat{n}(\mathbf{x}) | \Psi_{0}  \rangle
e^{i \mathbf{Q}\cdot(\mathbf{x} - \mathbf{x}^{\prime})}.
\end{eqnarray}
Here, $|\Psi_{0}\rangle$  and $E_{0}$ represent, respectively, the stationary electronic state (e.g., the ground state)
and the associated electronic energy.
The key quantity in Eq.~(\ref{eq5}) may be expressed in terms of the dynamic structure factor (DSF)
\begin{equation}\label{eq51}
S(\mathbf{Q}, \omega) = \sum_{f} \delta( \omega+ E_{0} - E_{f} )
\left |\int d^{3}x ~
\langle \Psi_{f} | \hat{n}(\mathbf{x}) | \Psi_{0}  \rangle
e^{i \mathbf{Q}\cdot\mathbf{x}} \right| ^{2},
\end{equation}
where $\omega =  \omega_{\mathbf{k}_{in}}- \omega_{\mathbf{k}_{s}}$ is the photon energy
transfer~\cite{schulke2007electron}. 
$S(\mathbf{Q},\omega)$ is the Fourier transform of the Van Hove
correlation function~\cite{van1954correlations}.
Note that if  the energy window function $W_{\Delta E}$ is centered at $\omega_{\mathbf{k}_{in}}$, and $\Delta E$ is small, 
such that $ \omega_{\mathbf{k}_{s}}= \omega_{\mathbf{k}_{in}}$, then 
 Eq.~(\ref{eq5}) reduces to elastic  x-ray scattering  from a stationary target.

In contrast to x-ray scattering from a stationary target, the pulse duration in TRXS determines the time resolution and has to be shorter than the motion of the electronic wave packet.
Therefore, in TRXS the incoming photon energy is not well defined, 
due to the inherent bandwidth of the x-ray pulse.
Comparing Eqs.~(\ref{eq441}) and (\ref{eq5}), one sees that TRXS is a generalized form of  
stationary-state x-ray scattering that depends on the spectrum of the pulse.
The generalized DSF for TRXS is 
\begin{equation}\label{eq63}
\tilde{S}(\mathbf{Q}, \omega, \tau_{d}) = \sum_{f}
\frac{\tau_{l}}{\sqrt{8 \pi \textrm{ln}2}} e^{
-\frac{\tau^{2}_{l}}{8 \textrm{ln}2} ( \omega+ \tilde{E} - E_{f})^{2}} \left | \int
d^{3}x ~ \langle \Psi_{f} | \hat{n}(\mathbf{x}, \tau_{d}) | \Phi
\rangle e^{i \mathbf{Q}\cdot\mathbf{x}} \right | ^{2}.
\end{equation}
In the following, we will discuss several experimental situations described by different spectral window functions, as well as
how the generalized DSF can be related to the linear density perturbation for 
electronic wave packets.

\subsection{Impact of energy resolution of scattering detector on TRXS}

In the following, we will consider the role of $W_{\Delta E}$ in
TRXS for two interesting situations for the energy resolution.

{\bf{i. No energy resolution}}

First, we consider the case where the detector does not resolve the energy of scattered photons.
In this case, the spectral window function can be treated as being constant and
$\Delta E$ as large enough to include all accessible scattered photon energies.
Here, we assume $\Delta E$ to be sufficiently large, but the object size $D$ to be sufficiently small,
such that the uniqueness of a pixel in $Q$-space, the pixel size being given by $\Delta Q = \pi /D$, is not lost due
to large uncertainty in the momentum distribution. 
For example, the Compton shift from a resting electron for x-rays with 10\,keV energy is about 
$\Delta \omega \approx 57\,\mathrm{eV}$ at a scattering angle of $45^{\circ}$. 
Thus assuming a maximum shift of about $2 \Delta \omega =114$ eV,
one finds a maximum object size of $D \approx 38\,\text{\AA}$.
In this situation the $\omega_{\mathbf{k}_{s}}$-dependent integral
[see Eq.~(\ref{eq441})] may be written as
\begin{equation}\label{eq61}
\int_{0}^{\infty} d\omega_{\mathbf{k}_{s}} \omega_{\mathbf{k}_{s}}
\frac{\tau_{l}}{\sqrt{8 \pi \textrm{ln}2}}
e^{-\frac{\tau^{2}_{l}}{8 \textrm{ln}2} ( \omega_{\mathbf{k}_{in}} - {\omega_{\mathbf{k}_{s}}}
+ \tilde{E} - E_{f})^{2}}
\simeq \omega_{\mathbf{k}_{in}}.
\end{equation}
Here, we assume that $\omega_{\mathbf{k}_{in}} \gg |E_{f} - \tilde{E}| $, i.e., 
due to the insufficient energy resolution the pulse can be treated as quasi-monochromatic.
Substituting the result from
Eq.~(\ref{eq61}) in Eq.~(\ref{eq441}), the expression for the DSP
in the case of no energy resolution becomes
\begin{align}\label{eq7}
\frac{dP}{d\Omega}  &=  \frac{d\sigma_{th}}{d\Omega}~\mathcal{F}~
\sum_{f} 
 \left |
\int d^{3}x ~ \langle \Psi_{f} | \hat{n}(\mathbf{x}, \tau_{d}) |
\Phi \rangle e^{i \mathbf{Q}\cdot\mathbf{x}} \right | ^{2} \\
  & = \frac{d\sigma_{th}}{d\Omega}~\mathcal{F}~
      \int d^{3}x \int d^{3}x' ~ 
      \langle \Phi| \hat{n}(\mathbf{x'}, \tau_{d}) \hat{n}(\mathbf{x}, \tau_{d}) | \Phi \rangle 
      e^{i \mathbf{Q}\cdot(\mathbf{x}-\mathbf{x'})} .
\end{align}
This result for the energy-integrated generalized DSF resembles the static case \cite{schulke2007electron}.
However, the DSP still depends on the pump-probe delay $\tau_d$.
The integrated DSF encodes the electron pair-correlation function \cite{van1954correlations}
and one could observe electron correlation effects in experiments \cite{petrillo, schuelke1995, watanabe}.
Thus, in the case of an electronic wave packet, the energy-integrated generalized DSF  offers the 
electron pair-correlation function of the wave packet at different delay times from which one can retrieve information 
about  time-dependent electronic correlations in the wave packet.
Therefore, wave packet dynamics can be imaged via TRXS even in the case of no energy resolution.

{\bf{ii. High energy resolution}}

Now, we consider the situation
where $\Delta E$ is much smaller than the bandwidth, i.e., high
energy resolution. 
Let the spectral window
function be centered at $\tilde{\omega}_{\mathbf{k}_{d}}$ and
\begin{equation}\label{eq6}
\int_{0}^{\infty} d\omega_{\mathbf{k}_{s}} W_{\Delta E} ({\omega_{\mathbf{k}_{s}}}) 
= \Delta E.
\end{equation} 
Hence, 
\begin{align}\label{eq62}
 & \int_{0}^{\infty} d\omega_{\mathbf{k}_{s}}
W_{\Delta E} ({\omega_{\mathbf{k}_{s}}}) 
\omega_{\mathbf{k}_{s}} \frac{\tau_{l}}{\sqrt{8 \pi \textrm{ln}2}}
e^{-\frac{\tau^{2}_{l}}{8 \textrm{ln}2} (
\omega_{\mathbf{k}_{in}} - {\omega_{\mathbf{k}_{s}}} + \tilde{E} - E_{f})^{2}} \nonumber \\
& \quad \simeq \tilde{\omega}_{\mathbf{k}_{d}} \Delta E
\frac{\tau_{l}}{\sqrt{8 \pi \textrm{ln}2}}
e^{-\frac{\tau^{2}_{l}}{8 \textrm{ln}2} ( {\omega_{\mathbf{k}_{in}} + \tilde{E} - E_{f}
-\tilde{\omega}_{\mathbf{k}_{d}})^{2}}} .
\end{align}
On substituting the result from Eq.~(\ref{eq62}) in
Eq.~(\ref{eq441}), the expression for the DSP in the case of high
energy resolution reduces to
\begin{equation}\label{eq8}
\frac{dP}{d\Omega}  =  \frac{d\sigma_{th}}{d\Omega} ~\mathcal{F}~
 \Delta E \sum_{f} \frac{\tau_{l}}{\sqrt{8
\pi \textrm{ln}2}} e^{-\frac{\tau^{2}_{l}}{8 \textrm{ln}2} ( {
\omega_{\mathbf{k}_{in}} + \tilde{E} - E_{f} -
\tilde{\omega}_{\mathbf{k}_{d}})^{2}}} \left | \int d^{3}x ~
\langle \Psi_{f} | \hat{n}(\mathbf{x}, \tau_{d}) | \Phi \rangle
e^{i \mathbf{Q}\cdot\mathbf{x}} \right | ^{2} ,
\end{equation}
assuming $\tilde{\omega}_{\mathbf{k}_{d}}/\omega_{\mathbf{k}_{in}} \simeq 1$.

Equation~(\ref{eq8}) shows that in the case of high energy resolution the DSP is determined by the generalized DSF  [see Eq.~(\ref{eq63})] for TRXS:
\begin{equation}\label{eq811}
\frac{dP}{d\Omega}  =  \frac{d\sigma_{th}}{d\Omega} ~\mathcal{F}~
 \Delta E ~\tilde{S}(\mathbf{Q}, \omega_{\mathbf{k}_{in}}-\tilde{\omega}_{\mathbf{k}_{d}}, \tau_{d}).
\end{equation}
Thus the scattering signal depends on the spectrum of the x-ray pulse and on the position of the window function $\tilde{\omega}_{\mathbf{k}_{d}}$.
The special case where the spectral window function is centered at the central frequency of the incoming beam, $\tilde{\omega}_{\mathbf{k}_{d}} = \omega_{\mathbf{k}_{in}}$,
comes closest to what can be considered time-resolved coherent diffractive imaging.
However, even then one does not just recover the Fourier transform of the instantaneous electron density:  
one measures the generalized DSF $\tilde{S}(\mathbf{Q},0, \tau_{d})$, which 
in contrast to the static case includes inelastic scattering within the bandwidth 
and can not be reduced to $|\int d^{3}x 
\langle \Phi | \hat{n}(\mathbf{x}, \tau_{d}) | \Phi \rangle
e^{i \mathbf{Q}\cdot\mathbf{x}}|^{2}$.
Thus, the scattering signal depends on the spatio-temporal density-density correlation function of the wave packet.
By changing  the position of the window function one can measure the generalized DSF $\tilde{S}(\mathbf{Q},\omega, \tau_{d})$. 
We probe the wave packet at different delay times $\tau_{d}$ with a time resolution given by the probe pulse duration $\tau_l$.
By measuring the generalized DSF it is possible to extract information about dynamics on time scales much faster than $\tau_l$.
To this end, in the next section we will combine TRXS of electronic wave packets with the approach of Abbamonte and co-workers.

\subsection{Linear response to density perturbations for electronic wave packets}

We showed in the last section that TRXS with high energy resolution depends on the generalized DSF. 
In this section we investigate which information about dynamics can be extracted from $\tilde{S}(\mathbf{Q}, \omega, \tau_{d})$.
In  x-ray scattering from a stationary, homogeneous target, $S$ is related to the propagator for the electron density $\chi$ by \cite{abbamonte2004imaging}
\begin{equation}\label{eq65}
\textrm{Im}\big[ \chi(\mathbf{Q}, \omega) \big]
 = - \pi [ S(\mathbf{Q}, \omega) - S(\mathbf{Q}, - \omega) ],
\end{equation}
i.e., the energy and momentum-resolved scattering signal is related
to the imaginary part of the electron density propagator in the Fourier domain. 
For a target in thermal equilibrium this is a version of the fluctuation-dissipation theorem~\cite{schulke2007electron}. 
In the real-space and real-time domain, 
$\chi(\mathbf{x} - \mathbf{x}',t - t')$ reflects the amplitude for some perturbation in the electron density to propagate from position $\mathbf{x}'$ to
$\mathbf{x}$ during a finite time interval $t-t'$~\cite{van1954correlations}. 
Experimental measurement only provides the imaginary part of $\chi$, via Eq.~(\ref{eq65}).
However, a four-step recipe to reconstruct the full $\chi (\mathbf{x}-\mathbf{x}', t-t')$ from
the experimentally accessible $\textrm{Im}[\chi(\mathbf{Q}, \omega)]$
has been developed by Abbamonte {\em et al.}~\cite{abbamonte2004imaging, abbamonte2010ultrafast}.
This approach has been applied to image
ultrafast electron dynamics at synchrotron light
sources~\cite{abbamonte2008dynamical, abbamonte2010ultrafast}.
The reconstructed $\chi (\mathbf{x}-\mathbf{x}', t-t')$ is the complete response for a homogeneous system. 
In the case of a stationary, inhomogeneous system \cite{abbamonte2009inhomogeneous} one recovers the propagator $\chi (\mathbf{x}, \mathbf{x}', t-t')$ averaged over all source locations $\mathbf{x}'$.
Recently, a method  based on a coherent standing wave source was proposed 
to obtain the full $\chi (\mathbf{x},\mathbf{x}', t-t')$ for inhomogeneous systems~\cite{abbamonte2012standingwaves}.

In our case of a non-stationary, inhomogeneous wave packet the x-ray scattering depends on the generalized DSF $\tilde{S}(\mathbf{Q},\omega,\tau_d)$.
Define a generalized electron density propagator 
\begin{equation}
 \tilde{\chi}(\mathbf{x},\mathbf{x}',t,t')= \chi(\mathbf{x},\mathbf{x}',t,t') C(t-t'),
\end{equation} 
connecting the density propagator $\chi$ and the temporal coherence function $C$  of the x-ray pulse, see Eq.~\eqref{eq35}.
Similar to Eq.~\eqref{eq65}, we find a relation between the generalized DSF and the generalized electron density propagator (see appendix~\ref{app:3}), 
\begin{align}
\mathrm{Im}\big[ \tilde{\chi}(\mathbf{Q},-\mathbf{Q}, \omega, \tau_{d}) \big] 
&= - \pi [ \tilde{S}(\mathbf{Q},\omega,\tau_d) - \tilde{S}(\mathbf{Q},-\omega,\tau_d) ]. 
\end{align}
Now, using the four-step recipe, as developed by Abbamonte {\em et al.}~\cite{abbamonte2004imaging, abbamonte2010ultrafast}, one can obtain the full $\tilde{\chi}(\mathbf{Q}, \omega, \tau_{d})$  from the experimentally accessible 
$\textrm{Im}[\tilde{\chi}(\mathbf{Q}, \omega,
\tau_{d})]$. In this way, the generalized electron density
propagator can be obtained.
Observe that when the temporal coherence function $C$ is known one obtains the exact propagator $\chi(\mathbf{x},\mathbf{x}',t,t')$ from $\tilde{\chi}(\mathbf{x},\mathbf{x}',t,t')$ by division and in any case the generalized propagator reduces to the exact propagator if $t-t'$ is much shorter than the pulse duration.

In the last section, we have seen that TRXS from a wave packet is complicated by electron density dynamics faster than the pulse duration, see Eq.~(\ref{eq43}).
Therefore, it is natural to ask the question whether the electron density propagator obtained from $\tilde{S}(\mathbf{Q}, \omega, \tau_{d})$ can be used to unravel these induced dynamics.
To answer this we investigate the linear density response of an electronic wave packet to the scattering process.
Note that here we analyze the response of the exact propagator $\chi(\mathbf{x},\mathbf{x}', t, t')$.
The detailed derivation is given in appendix~\ref{app:2}.
The true physical density response can be written as
\begin{equation}\label{eqa001}
\delta{n}(\mathbf{x}, t) = \textrm{Tr}[\hat{n}(\mathbf{x}, t) \delta \hat{\rho}(t)],
\end{equation}
where $\delta \hat{\rho}(t)$  is the change in the electronic state within linear response theory. 
For $\mathbf{\hat{A}^{2}}$-induced non-resonant scattering, the linear-order terms of the above equation can be written as
\begin{equation}\label{eqa008}
\delta{n}(\mathbf{x}, t) = (-i)  \frac{\alpha^{2}}{2} 
\int_{-\infty}^{t} dt^{\prime} \int d^{3}x^{\prime}
~\textrm{Tr}[\hat{\rho}^{X}_{in}  \hat{\mathbf{A}}^{2}(\mathbf{x}^{\prime}, t^{\prime})]
~\langle \Phi | [\hat{n}(\mathbf{x}, t),
\hat{n}(\mathbf{x}^{\prime}, t^{\prime})] | \Phi \rangle. 
\end{equation}
Here, $ \hat{\rho}^{X}_{in}$ represents the initial density operator for
the x rays.  

Performing some simple mathematical steps (see appendix~\ref{app:2})
one obtains the linear density response of $\mathbf{\hat{A}^{2}}$ scattering.
The contributions from photon scattering expressed by the field correlation functions
$\langle \mathbf{\hat{E}}^{(\pm)}(\mathbf{x}^{\prime}, t^{\prime})
\mathbf{\hat{E}}^{(\mp)}(\mathbf{x}^{\prime}, t^{\prime}) \rangle$ to
the linear density response are zero. 
The only non-zero contributions to the linear density response come from the field correlation functions 
$\langle \mathbf{\hat{E}}^{(\pm)}(\mathbf{x}^{\prime}, t^{\prime})
\mathbf{\hat{E}}^{(\pm)}(\mathbf{x}^{\prime}, t^{\prime}) \rangle$ when the field has a fixed carrier envelope phase.
In typical experiments, however, the phase is not controlled and one has to average over the phase even in our ideal case of a Gaussian ensemble.
Thus, the linear density response of the $\mathbf{\hat{A}^{2}}$ scattering process itself vanishes.
Therefore, the fast dynamics induced in $\mathbf{\hat{A}^{2}}$ non-resonant scattering cannot be captured by the linear response electron density propagator $\chi$.
This does not render $\chi$ meaningless.
Our finding rather expresses the fact that $\chi$ describes linear-order density fluctuations, whereas the density response to non-resonant x-ray scattering is, in general, a higher-order process.

\subsection{TRXS from a crystal: recovery of instantaneous electron density}

In this final subsection, we consider the case of TRXS from a crystal of identical electronic wave packets.
We separate the coherent and the incoherent scattering by inserting a complete set of energy eigenstates that is projected onto
the initial state and its orthogonal complement 
\begin{equation}
 |\Phi\rangle  \langle \Phi| + (1-|\Phi\rangle  \langle \Phi|)
	= |\Phi\rangle  \langle \Phi| + (1-|\Phi\rangle  \langle \Phi|) \sum_f |\Psi_f \rangle  \langle \Psi_f| (1-|\Phi\rangle  \langle \Phi|) .  
\end{equation}
Now we insert this complete set into Eq.~(\ref{eq43}).
As in section~\ref{secII} we assume a pulse short enough to freeze the wave packet motion.
In particular, we exploit that $| E_i - \tilde{E}| \ll 1/\delta$ for eigenstates $|\Psi_i \rangle$ contained in the wave packet with energies $E_i$ and thus
$$ e^{i\hat{H}\delta} (1 - |\Phi\rangle\langle\Phi| ) |\Psi_i\rangle 
\approx e^{i E_i\delta}|\Psi_i\rangle -  e^{i \tilde{E}\delta} |\Phi\rangle\langle\Phi|\Psi_i\rangle
\approx e^{i E_i\delta} (1 - |\Phi\rangle\langle\Phi| ) |\Psi_i\rangle.$$
The key expression of Eq.~(\ref{eq44}) can then be rewritten
\begin{align}\label{eqa42}
&\int d^{3}x \int d^{3}x^{\prime} 
\biggl[
e^{ -\frac{\tau^{2}_{l}}{8 \textrm{ln}2} ( \omega_{\mathbf{k}_{in}}- {\omega_{\mathbf{k}_{s}}})^2 }
 \langle \Phi| \hat{n}(\mathbf{x}^{\prime}, \tau_{d})
| \Phi \rangle \langle \Phi| \hat{n}(\mathbf{x}, \tau_{d})| \Phi \rangle \nonumber\\
&\qquad +
\sum_f 
e^{ -\frac{\tau^{2}_{l}}{8 \textrm{ln}2} ( \omega_{\mathbf{k}_{in}}- {\omega_{\mathbf{k}_{s}}} + \tilde{E} - E_f )^2 }
\langle \Phi| \hat{n}(\mathbf{x}^{\prime}, \tau_{d})
| \Psi_f' \rangle \langle \Psi_f'| \hat{n}(\mathbf{x}, \tau_{d})| \Phi \rangle
\biggr] 
e^{i\mathbf{Q}\cdot(\mathbf{x} - \mathbf{x}^{\prime})} \nonumber\\ 
&= 
e^{ -\frac{\tau^{2}_{l}}{8 \textrm{ln}2} ( \omega_{\mathbf{k}_{in}}- {\omega_{\mathbf{k}_{s}}} )^2 }
\left| \int d^{3}x \,  \langle \Phi| \hat{n}(\mathbf{x}, \tau_{d})| \Phi \rangle \, e^{i\mathbf{Q}\cdot\mathbf{x}} \right|^2 
+
\sum_f 
e^{ -\frac{\tau^{2}_{l}}{8 \textrm{ln}2} ( \omega_{\mathbf{k}_{in}}- {\omega_{\mathbf{k}_{s}}} + \tilde{E} - E_f )^2 } \nonumber\\
&\qquad \times 
\int d^{3}x \int d^{3}x^{\prime}
\langle \Phi| \hat{n}(\mathbf{x}^{\prime}, \tau_{d})
| \Psi_f' \rangle \langle \Psi_f'| \hat{n}(\mathbf{x}, \tau_{d})| \Phi \rangle
e^{i\mathbf{Q}\cdot(\mathbf{x} - \mathbf{x}^{\prime})}
\;,
\end{align}
where $|\Psi_f'\rangle = (1-|\Phi\rangle  \langle \Phi|) |\Psi_f \rangle$.

\begin{figure}[ht]
\includegraphics[width=6cm]{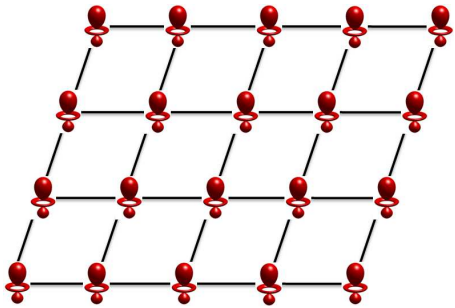}
\caption{(color online). A two dimensional view of a crystal made
of identical atoms prepared in exactly the same quantum
superposition and with identical phase.} \label{fig2}
\end{figure}

Now consider a crystal where an identical electronic wave packet is prepared at each lattice site with the help
of a pump pulse (see Fig.~\ref{fig2}). 
We assume the sub-units of the crystal to be noninteracting.
The electronic states in Eq.~(\ref{eqa42}) represent the state of the entire crystal, which factorizes into 
the electronic states of the individual sub-units.
Due to the periodic structure of the crystal, the first term in Eq.~(\ref{eqa42})
provides a coherent scattering signal giving rise to 
Bragg reflections.
According to the Laue condition, for a sufficiently large crystal, the lattice sum allows scattering only at momentum transfer $\mathbf{Q}$ 
that is equal to a reciprocal lattice vector. 
For the coherent scattering, the lattice sum is a coherent sum, because it is impossible to distinguish at which sub-unit the scattering occurred.
Thus, the Bragg intensity of the coherent scattering signal scales with the square of the number of unit cells
in the crystal. 
The second term in Eq.~(\ref{eqa42}) describes an incoherent scattering signal,
where at one lattice site an electronic transition from the wave packet 
to a state that is  not  part of the wave packet is induced.
Therefore, the sum over final states in Eq.~(\ref{eqa42}) involves a sum over the different lattice sites.
Because the site where the wave packet was destroyed can be distinguished from the other sites, 
the corresponding contributions must be summed incoherently. 
Thus, the intensity of the incoherent signal scales only linearly with the number of unit cells in
the crystal.

In conclusion, the TRXS signal is dominated by the coherent scattering
signal for a sufficiently large crystal.
In the case of TRXS from a crystal, the instantaneous electron density 
$\langle \Phi | \hat{n}(\mathbf{x}, \tau_{d}) | \Phi \rangle$
of the wave
packet can be retrieved from the coherent scattering signal. 
Although for a short pulse the bandwidth is large, the coherent signal dominates in the case of sufficiently many unit cells.
It is important to mention that in the case of scattering from a single electronic wave packet,
as demonstrated in our previous work~\cite{dixit2012, dixit2013jcp},
the contribution from the incoherent scattering signal dominates over the contribution from the 
coherent scattering signal.

\section{Conclusion}
This work is devoted to a rigorous understanding of TRXS to image 
coherent electronic motion on an ultrafast timescale using ultrashort hard x-ray pulses. 
The role of  the pulse duration and of the energy resolution of the scattering detector have been investigated  
for TRXS. 
For stationary targets, long probe pulses can be used and the theory reduces to non-resonant 
x-ray scattering (elastic and inelastic).
To image electronic wave packets one has to use ultrashort x-ray pulses.
It is found that TRXS encodes the generalized dynamic structure factor and not the instantaneous electron density of an electronic wave packet.
We have analyzed the scattering signal for two limiting situations of the energy resolution of the scattering detector.
In both situations, the probe pulse duration sets the time resolution for the electronic motion. 
In the case of no energy resolution the scattering signal depends on the wave packet dynamics through the pump-probe delay and
one measures the energy-integrated DSF which encodes the time-dependent electron pair-correlation function of the wave packet.
Therefore, TRXS with no energy resolution is of particular importance as it seems feasible with existing detector technology 
and offers to image time-dependent electron correlations in dynamical electron systems.
In the case of high energy resolution (small $\Delta E$) the scattering signal contains additional fingerprints of dynamics 
induced by the scattering process that are faster than the probe duration.
They can be probed in the energy domain by varying the position
($\tilde{\omega}_{\mathbf{k}_{d}}$) of the
spectral window function.
We have made the connection of our theory to the dynamical imaging of ultrafast dynamics by Abbamonte and co-workers.
In this way we have shown that from energy resolved x-ray scattering one can recover the electron density propagator for time scales much faster than the probe pulse duration. 
Thus, the present theory can be regarded as a unified description of time-resolved ultrafast x-ray scattering.  
The response of the density perturbation for an electronic wave packet
within the linear-response theory has been investigated. We showed
that the linear density response due to the scattering event itself vanishes.
A special and interesting situation has been considered for TRXS from a crystal, consisting of identical electronic wave packets 
with identical phase at each lattice point.
The scattering signal of the crystal is shown to recover the instantaneous electron density of the wave
packet in this case.  We hope that our present analysis of TRXS on ultrafast timescales will help in planning 
and understanding future experiments.

\appendix
\section{First-order correlation function for x rays}\label{app:1}
The first-order correlation function is defined as
\begin{eqnarray}\label{eq31}
G^{(1)} (\mathbf{x}_{1}, t_{1}; \mathbf{x}_{2}, t_{2}) & = &
\textrm{Tr}[\hat{\rho}^{X}_{in} \hat{E}^{(-)}(\mathbf{x}_{1}, t_{1}) \hat{E}^{(+)}(\mathbf{x}_{2}, t_{2})] \nonumber \\
& = & \langle E^{(-)}(\mathbf{x}_{1}, t_{1}) E^{(+)}(\mathbf{x}_{2}, t_{2})  \rangle,
\end{eqnarray}
where $\hat{\rho}^{X}_{in}$ is the initial density operator for
the x rays and $\hat{E}^{(+)} (\hat{E}^{(-)})$ is the positive
(negative) component of the electric field operator. In the
classical limit, the pulsed electric field can be expressed as
\begin{equation}\label{eq32}
E(\mathbf{x}, t) = E_{0}~ \textrm{cos}(\mathbf{k}_{in} \cdot
\mathbf{x} - \omega_{in} [t-\tau_{d}])
~g(\mathbf{k}_{in} \cdot \mathbf{x} - \omega_{in}
[t-\tau_{d}]),
\end{equation}
where $\mathbf{k}_{in}$ and $\omega_{in}$ are, respectively,  the
carrier wave vector and the frequency of the pulsed field.
Here, we assume that the envelope function $g$ is
Gaussian:
\begin{equation}\label{eq33}
g(\mathbf{k}_{in} \cdot \mathbf{x} - \omega_{in}
[t-\tau_{d}]) =  
e^{-\frac{1}{2 \sigma^{2}\omega^2_{in} } 
(\mathbf{k}_{in} \cdot \mathbf{x} -
\omega_{in} [t-\tau_{d}])^{2} },
\end{equation}
where $\tau_{d}$ is the time delay and $\sigma$ is related to the pulse duration as $\tau_{l} = \sqrt{8\ln 2}~\sigma$. 
The probe pulse is assumed to
propagate along the $z$ direction. Therefore, by using the
Eqs.~(\ref{eq32}) and (\ref{eq33}), the key quantity for the
first-order correlation function is expressed as
\begin{eqnarray}\label{eq34}
E^{(-)}(z_{1}, t_{1}) E^{(+)}(z_{2}, t_{2}) & = &
\frac{1}{4}|E_{0}|^{2}~ e^{-\frac{1}{2 \sigma^{2}} ( \alpha^2
z_{1}^{2} + [t_{1}-\tau_{d}]^{2} - 2
\alpha z_{1} [t_{1}-\tau_{d}])}
e^{i(k_{in} z_{1} - \omega_{in} t_{1} )} \nonumber \\
& & \times  e^{-\frac{1}{2 \sigma^{2}} (\alpha^2 z_{2}^{2} +
[t_2-\tau_{d}]^{2} - 2 \alpha z_{2}
[t_{2}-\tau_{d}])} e^{-i(k_{in} z_{2} -
\omega_{in} t_{2})}.
\end{eqnarray}
At this point, it is important to analyze the spatial dependence
in Eq.~(\ref{eq34}), i.e., under which conditions the spatially
dependent terms can be ignored and all the electrons in the sample
experience a spatially uniform pulse envelope. \ The spatially
dependent terms, $\textrm{exp}[- \alpha^2 z^{2}/ 2\sigma^{2}]$
and $\textrm{exp}[ \alpha z t /
\sigma^{2}]$, can be approximated by unity if the condition 
$\alpha |z| \ll\tau_{l}$ is satisfied. Here, $\tau_{l}$ is the pulse
duration and $1/\alpha$ is the speed of light. Let us consider the
situation considered in our previous works~\cite{dixit2012,
dixit2013jcp}, where a 1-fs pulse duration has been used.
Therefore, for the given value of $\tau_{l}$, $ |z| \ll \tau_{l}/\alpha
\simeq 300$~ nanometers. Hence, for a sample size of tens of nanometers (or
smaller) exposed to a few-fs hard x-ray pulse,
the spatial dependence of the envelope of the incident pulse can
be ignored.
Note that we assume the object size to be small in comparison to the transverse size of the x-ray beam.
Therefore, the first-order correlation function for x
rays can be written as
\begin{equation}\label{eq35}
G^{(1)} (\mathbf{x}_{1}, t_{1}; \mathbf{x}_{2}, t_{2}) = 2 \pi
\alpha I(\tau) C(\delta) e^{-i \omega_{in} \delta}  
e^{i\mathbf{k}_{in} \cdot ( \mathbf{x} - \mathbf{x}^{\prime} )},
\end{equation}
where $\tau = \frac{t_{1}+t_{2}}{2}$, $\delta = t_{2}-t_{1}$, and $C(\delta)= e^{- 2 \ln{2}\; \delta^{2}/\tau_{l}^{2}}$.
Observe that the function $C(\delta)$ 
and thus the first-order correlation function vanish for time differences $\delta$ much larger than the pulse duration.
Therefore, the function $C(\delta)$ describes the temporal coherence properties of the x rays.

\section{Density perturbation for an electronic wave packet  within linear response theory }\label{app:2}

The physical density response can be written as
\begin{equation}\label{eqa01}
\delta{n}(\mathbf{x}, t) = \textrm{Tr}[\hat{n}(\mathbf{x}, t) \delta \hat{\rho}(t)],
\end{equation}
where $ \delta \hat{\rho}(t)$ is the change in the density operator within linear-response theory and can be written as
\begin{equation}\label{eqa05}
\delta \hat{\rho}(t)  = \hat{\rho}^{(0,1)}(t) + \hat{\rho}^{(1,0)}(t),
\end{equation}
with
\begin{equation}\label{eqa06}
\hat{\rho}^{(0,1)}(t) =  \sum_{\{n\},\{\bar{n}\}} \rho^{X}_{\{n\},\{\bar{n}\}}
|\Psi_{\{n\}} \rangle \langle \Psi_{\{\bar{n}\}}^{(1)}, t|.
\end{equation}
Here, $\rho^{X}_{\{n\},\{\bar{n}\}}$ represents the populations and coherences of all the occupied field modes of the incident radiation.
Here, we have assumed that with the help of a pump pulse an electronic wave packet, $|\Phi \rangle$, is prepared and
$|\Psi_{\{n\}} \rangle = |\Phi \rangle | \{n\} \rangle$ is a product state of electronic and photon states.
The first order change of the state vector in the interaction picture can be written as
\begin{equation}\label{eqa04}
|\Psi^{(1)}_{\{n\}}, t \rangle = (-i) \int_{-\infty}^{t}
dt^{\prime} \hat{H}_{\textrm{int}}(t^{\prime})
|\Psi_{\{n\}}\rangle.
\end{equation}
On substituting the expressions from Eqs.~(\ref{eqa05}) and ~(\ref{eqa06}) in Eq.~(\ref{eqa01}),
the physical density-response can be expressed as
\begin{eqnarray}\label{eqa07}
\delta{n}(\mathbf{x}, t) & = &  \textrm{Tr}[\hat{n}(\mathbf{x}, t) (\hat{\rho}^{(0,1)}(t) + \hat{\rho}^{(1,0)}(t))] \nonumber \\
& = & -i \sum_{\{n\},\{\bar{n}\}} \rho^{X}_{\{n\},\{\bar{n}\}} \int_{-\infty}^{t} dt^{\prime}
\langle \Psi_{\{\bar{n}\}} |
[\hat{n}(\mathbf{x}, t), \hat{H}_{\textrm{int}}(t^{\prime})]  |\Psi_ {\{n\}} \rangle.
\end{eqnarray}
Now, on using the second term of $\hat{H}_{\textrm{int}}$ as shown in Eq.~(\ref{eq1}) ($\hat{\mathbf{A}}^{2}
$ term), the above equation can be written as
\begin{equation}\label{eqa08}
\delta{n}(\mathbf{x}, t) = (-i)  \frac{\alpha^{2}}{2} 
\int_{-\infty}^{t} dt^{\prime} \int d^{3}x^{\prime}
\textrm{Tr}[\hat{\rho}^{X}_{in}  \hat{\mathbf{A}}^{2}(\mathbf{x}^{\prime}, t^{\prime})]
\langle \Phi | [\hat{n}(\mathbf{x}, t),
\hat{n}(\mathbf{x}^{\prime}, t^{\prime})] | \Phi \rangle, 
\end{equation}
which can be expressed in the following known form as~\cite{fetter1971}
\begin{equation}\label{eqa09}
\delta{n}(\mathbf{x}, t) =
\int_{-\infty}^{t} dt^{\prime} \int d^{3}x^{\prime}
\chi(\mathbf{x}, \mathbf{x}^{\prime}, t, t^{\prime}) V(\mathbf{x}^{\prime}, t^{\prime}).
\end{equation}
On comparing Eq.~(\ref{eqa08}) with Eq.~(\ref{eqa09}), we can write the linear-response function
at position $\mathbf{x}$ and time $t$ due to the external potential at position $\mathbf{x}^{\prime}$
and time $t^{\prime}$ as
\begin{equation}\label{eqa10}
\chi(\mathbf{x}, \mathbf{x}^{\prime}, t, t^{\prime}) = -i ~ \langle \Phi | [\hat{n}(\mathbf{x}, t),
\hat{n}(\mathbf{x}^{\prime}, t^{\prime})] | \Phi \rangle,
\end{equation}
and the interaction potential as
\begin{equation}\label{eqa11}
V(\mathbf{x}^{\prime}, t^{\prime}) = \frac{\alpha^{2}}{2} 
\textrm{Tr}[\hat{\rho}^{X}_{in}  \hat{\mathbf{A}}^{2}(\mathbf{x}^{\prime}, t^{\prime})]. 
\end{equation}
In the following, we will simplify $V(\mathbf{x}^{\prime}, t^{\prime})$, which can be expressed in terms
of the electric field operator as
\begin{eqnarray}\label{eqa12}
V(\mathbf{x}^{\prime}, t^{\prime}) & = &  \frac{\alpha^{2}}{2 \omega_{\mathbf{k}_{in}}^{2}} 
\textrm{Tr}[\hat{\rho}^{X}_{in}  \hat{\mathbf{E}}^{2}(\mathbf{x}^{\prime}, t^{\prime})]  \nonumber \\
& = & \frac{\alpha^{2}}{2 \omega_{\mathbf{k}_{in}}^{2}} \left[
\textrm{Tr}[\hat{\rho}_{in}^{X} \hat{\mathbf{E}}^{(-)}(\mathbf{x}^{\prime}, t^{\prime})
\hat{\mathbf{E}}^{(+)}(\mathbf{x}^{\prime}, t^{\prime})] +
\textrm{Tr}[\hat{\rho}_{in}^{X} \hat{\mathbf{E}}^{(+)}(\mathbf{x}^{\prime}, t^{\prime})
\hat{\mathbf{E}}^{(-)}(\mathbf{x}^{\prime}, t^{\prime})] \right . \nonumber \\
&& \left. + \textrm{Tr}[\hat{\rho}_{in}^{X} \hat{\mathbf{E}}^{(+)}(\mathbf{x}^{\prime}, t^{\prime})
\hat{\mathbf{E}}^{(+)}(\mathbf{x}^{\prime}, t^{\prime})]
+ \textrm{Tr}[\hat{\rho}_{in}^{X} \hat{\mathbf{E}}^{(-)}(\mathbf{x}^{\prime}, t^{\prime})
\hat{\mathbf{E}}^{(-)}(\mathbf{x}^{\prime}, t^{\prime})] \right ].
\end{eqnarray}
On using the expressions for the pulsed electric field and envelope
function [see Eqs.~(\ref{eq32}) and (\ref{eq33}), respectively],
and using a similar procedure as shown in
appendix~(\ref{app:1}), the above equation can be simplified as
\begin{equation}\label{eqa13}
V(\mathbf{x}^{\prime}, t^{\prime})
= \frac{\alpha^{2}}{2\omega_{\mathbf{k}_{in}}^{2}}
\left[ \frac{1}{2} E_{0} g(\mathbf{k}_{in} \cdot \mathbf{x}^{\prime} - \omega_{\mathbf{k}_{in}} [t^{\prime}-\tau_{d}]) \right]^{2}
\left[ 2+ e^{-2i(\mathbf{k}_{in} \cdot \mathbf{x}^{\prime} - \omega_{\mathbf{k}_{in}} t^{\prime})} +
e^{2i(\mathbf{k}_{in} \cdot \mathbf{x}^{\prime} - \omega_{\mathbf{k}_{in}} t^{\prime})} \right].
\end{equation}
Here, we assume that all the electrons in the sample experience a uniform pulse envelope 
and hence the spatial dependency of the envelope function
can be ignored [see appendix~(\ref{app:1})]. Therefore, Eq.~(\ref{eqa13}) reduces to
\begin{equation}\label{eqa14}
V(\mathbf{x}^{\prime}, t^{\prime})
= \frac{\pi \alpha^{3}}{\omega_{\mathbf{k}_{in}}^{2}} I(t^{\prime})
\left[ 2+ e^{-2i(\mathbf{k}_{in} \cdot \mathbf{x}^{\prime} - \omega_{\mathbf{k}_{in}} t^{\prime})} +
e^{2i(\mathbf{k}_{in} \cdot \mathbf{x}^{\prime} - \omega_{\mathbf{k}_{in}} t^{\prime})} \right].
\end{equation}
Therefore, on substituting the expressions from Eqs.~(\ref{eqa13}) and ~(\ref{eqa10}), the density response can be written as
\begin{equation}\label{eqa15}
\delta{n}(\mathbf{x}, t) =
\frac{2 \pi \alpha^{3}}{\omega_{\mathbf{k}_{in}}^{2}}
\int_{-\infty}^{t} dt^{\prime} I(t^{\prime}) \int d^{3}x^{\prime}
\chi(\mathbf{x}, \mathbf{x}^{\prime}, t, t^{\prime})
\left[ 1+ \textrm{cos}[2(\mathbf{k}_{in} \cdot \mathbf{x}^{\prime} - \omega_{\mathbf{k}_{in}} t^{\prime})]
\right].
\end{equation}
The total density response can be decomposed into two parts as
\begin{equation}\label{eqa151}
\delta{n}(\mathbf{x}, t) = \delta{n}_{1}(\mathbf{x}, t) + \delta{n}_{2}(\mathbf{x}, t),
\end{equation}
where
\begin{equation}\label{eqa152}
\delta{n}_{1}(\mathbf{x}, t) = \frac{2\pi \alpha^{3}}{\omega_{\mathbf{k}_{in}}^{2}}
\int_{-\infty}^{t} dt^{\prime} I(t^{\prime}) \int d^{3}x^{\prime}
\chi(\mathbf{x}, \mathbf{x}^{\prime}, t, t^{\prime}),
\end{equation}
and
\begin{equation}\label{eqa153}
\delta{n}_{2}(\mathbf{x}, t) = \frac{2 \pi \alpha^{3}}{\omega_{\mathbf{k}_{in}}^{2}}
\int_{-\infty}^{t} dt^{\prime} I(t^{\prime}) \int d^{3}x^{\prime}
\chi(\mathbf{x}, \mathbf{x}^{\prime}, t, t^{\prime})
\textrm{cos}[2(\mathbf{k}_{in} \cdot \mathbf{x}^{\prime} - \omega_{\mathbf{k}_{in}} t^{\prime})]. 
\end{equation}
On performing the
$\mathbf{x}^{\prime}$-dependent integral in Eq.~(\ref{eqa152}), one
finds that $\delta{n}_{1}(\mathbf{x}, t) = 0$. 
Thus, for perfectly stable (coherent) x rays, 
the density response
for an electronic wave packet can be written as
\begin{equation}\label{eqa16}
\delta{n}(\mathbf{x}, t)
 =  \frac{2 \pi \alpha^{3}}{\omega_{\mathbf{k}_{in}}^{2}}
\int_{-\infty}^{t} dt^{\prime} I(t^{\prime}) \int d^{3}x^{\prime}
\chi(\mathbf{x}, \mathbf{x}^{\prime}, t, t^{\prime}) 
\textrm{cos}[2(\mathbf{k}_{in} \cdot \mathbf{x}^{\prime} - \omega_{\mathbf{k}_{in}} t^{\prime})]. 
\end{equation}

\section{Connection between linear-response function and dynamical structure factor}\label{app:3}

The retarded electron density propagator
\begin{equation}\label{eqC1}
\chi(\mathbf{x}, \mathbf{x}^{\prime}, t, t^{\prime}) = -i \, \langle \Phi | [\hat{n}(\mathbf{x}, t),
\hat{n}(\mathbf{x}^{\prime}, t^{\prime})] | \Phi \rangle \, \theta(t-t'),
\end{equation}
describes the  propagation of disturbances in the electron density and characterizes the electronic system.
The step function $\theta(t-t')$ ensures causality.
To connect the dynamical properties of the system with the temporal coherence of the probe pulse, we define 
the generalized propagator
\begin{equation}\label{eqc2}
\tilde{\chi}(\mathbf{x}, \mathbf{x}^{\prime}, t, t^{\prime}) = -i \,C(t-t')\, \langle \Phi | [\hat{n}(\mathbf{x}, t),
\hat{n}(\mathbf{x}^{\prime}, t^{\prime})] | \Phi \rangle \, \theta(t-t') ,
\end{equation}
where $C(t-t') = \exp[{-2\ln{2}\;(t-t')^{2}/\tau_{l}^{2}}]$ describes the temporal coherence of the x-ray pulse with pulse duration $\tau_l$, see Eq.~\eqref{eq35}. 
Observe that this propagator vanishes when $t-t'$ is much larger than the pulse duration.
As before, we image the system at a pump-probe delay time $\tau_d$ with a pulse duration short enough to freeze the wave packet.
Thus, $|t-\tau_d|$ and $|t'-\tau_d|$ of interest are small with respect to the wave packet motion and
we can write the generalized propagator in terms of $\delta=t-t'$ and $\tau_d$,
\begin{eqnarray}\label{eqa21}
\tilde{\chi}(\mathbf{x}, \mathbf{x}^{\prime}, \delta, \tau_d)
&:=&
\tilde{\chi}(\mathbf{x}, \mathbf{x}^{\prime}, t, t^{\prime}) \\
& = & -i \; C(\delta) \, \theta(\delta) 
\sum_{f}\left[ \langle \Phi | \hat{n}(\mathbf{x},\tau_d) | \Psi_{f} \rangle \langle \Psi_{f} |
\hat{n}(\mathbf{x}^{\prime},\tau_d) | \Phi \rangle   e^{i(\tilde{E}-E_{f})\delta}
 \right. \nonumber \\
&& \left. - \langle \Phi | \hat{n}(\mathbf{x}^{\prime},\tau_d) | \Psi_{f} \rangle
\langle \Psi_{f} | \hat{n}(\mathbf{x},\tau_d) | \Phi \rangle
e^{i(E_f-\tilde{E})\delta}
 \right] .
\end{eqnarray}
The Fourier transform with respect to $ \mathbf{x}, \mathbf{x}'$ and $\delta$ is given by 
\begin{align}\label{eqC3}
\tilde{\chi}(\mathbf{Q}, \mathbf{Q}', \omega, \tau_{d}) 
&=
\iiint d\delta d^3 x\, d^3 x'
e^{i\omega \delta}
e^{i \mathbf{Q} \cdot \mathbf{x}} \,
e^{i \mathbf{Q}' \cdot \mathbf{x}'} \;
\tilde{\chi}(\mathbf{x}, \mathbf{x}', \delta, \tau_{d})\\
&= 
-i \sum_{f}
\iint d^3 x\, d^3 x' e^{i (\mathbf{Q} \cdot \mathbf{x}+\mathbf{Q}' \cdot \mathbf{x}')} 
\int d\delta \, C(\delta) \, \theta(\delta) \notag\\
& \quad 
\times \Big[ e^{i (\omega+\tilde{E}-E_f) \delta}
\langle \Phi | \hat{n}(\mathbf{x},\tau_d) | \Psi_{f} \rangle \langle \Psi_{f} |
\hat{n}(\mathbf{x}',\tau_d) | \Phi \rangle \notag\\
&\quad - 
e^{i (\omega-\tilde{E}+E_f) \delta}
\langle \Phi | \hat{n}(\mathbf{x}',\tau_d) | \Psi_{f} \rangle \langle \Psi_{f} |
\hat{n}(\mathbf{x},\tau_d) | \Phi \rangle
 \Big].
\end{align}
Now, one can easily determine the imaginary part of $\tilde{\chi}(\mathbf{Q}, -\mathbf{Q}, \omega, \tau_{d})$ 
\begin{align}
\mathrm{Im}\big[ \tilde{\chi}(\mathbf{Q}, -\mathbf{Q}, \omega, \tau_{d}) \big] 
&= -\pi \sum_f \frac{\tau_l}{\sqrt{8\pi \ln 2}} e^{- \frac{\tau_l}{8\ln 2} (\omega +\tilde{E} -E_f)^2}  
	\left| \int d^3 x\, \langle \Psi_f | \hat{n}(\mathbf{x},\tau_d) | \Phi\rangle e^{i \mathbf{Q} \cdot \mathbf{x}} \right|^2 \notag\\
&\quad + \pi \sum_f \frac{\tau_l}{\sqrt{8\pi \ln 2}} e^{- \frac{\tau_l}{8\ln 2} (-\omega +\tilde{E} -E_f)^2}  
	\left| \int d^3 x\, \langle \Psi_f | \hat{n}(\mathbf{x},\tau_d) | \Phi\rangle e^{i \mathbf{Q} \cdot \mathbf{x}} \right|^2 \\
&= - \pi [ \tilde{S}(\mathbf{Q},\omega,\tau_d) - \tilde{S}(\mathbf{Q},-\omega,\tau_d) ]. 
\end{align}
This establishes the relation of the measured generalized DSF and the imaginary part of the generalized density propagator.
Applying the four-step recipe by Abbamonte and coworkers \cite{abbamonte2004imaging, abbamonte2010ultrafast} one can reconstruct real-space information about $\tilde{\chi}$.
From the definition of the generalized propagator, we see that
for time propagation much shorter than the pulse duaration ( $\delta\ll \tau_l$) this gives information about 
the density propagator.
Observe, that the generalized DSF only provides the diagonal terms, where $\mathbf{Q}'=-\mathbf{Q}$.
It was shown in Ref. \cite{abbamonte2009inhomogeneous} that one recovers the full electron density propagator only for homogeneous systems, whereas for the case of an inhomogeneous system  one obtains the averaged generalized propagator 
$$\tilde{\chi}(\mathbf{x},\delta,\tau_d) = \int d^3 x'\, \tilde{\chi}(\mathbf{x}+\mathbf{x}',\mathbf{x}',\delta,\tau_d)\,,$$ 
averaged over all possible source locations $\mathbf{x}'$.

%\bibliography{imaging} 

\begin{thebibliography}{10}

\bibitem{Ihee}
H.~Ihee, M.~Lorenc, T.~K. Kim, Q.~Y. Kong, M.~Cammarata, J.~H. Lee, S.~Bratos,
  and M.~Wulff,
\newblock Science {\bf 309}, 1223 (2005).

\bibitem{Chapman}
H.~N. Chapman, P.~Fromme, A.~Barty, T.~A. White, R.~A. Kirian, A.~Aquila, M.~S.
  Hunter, J.~Schulz, D.~P. DePonte, U.~Weierstall, R.~B. Doak, F.~R. N.~C.
  Maia, A.~V. Martin, I.~Schlichting, L.~Lomb, N.~Coppola, R.~L. Shoeman, S.~W.
  Epp, R.~Hartmann, D.~Rolles, A.~Rudenko, L.~Foucar, N.~Kimmel,
  G.~Weidenspointner, P.~Holl, M.~Liang, M.~Barthelmess, C.~Caleman, S.~Boutet,
  M.~J. Bogan, J.~Krzywinski, C.~Bostedt, S.~Bajt, L.~Gumprecht, B.~Rudek,
  B.~Erk, C.~Schmidt, A.~Hömke, C.~Reich, D.~Pietschner, L.~Strüder, G.~Hauser,
  H.~Gorke, J.~Ullrich, S.~Herrmann, G.~Schaller, F.~Schopper, H.~Soltau, K.~U.
  Kühnel, M.~Messerschmidt, J.~D. Bozek, S.~P. Hau-Riege, M.~Frank, C.~Y.
  Hampton, R.~G. Sierra, D.~Starodub, G.~J. Williams, J.~Hajdu, N.~Timneanu,
  M.~M. Seibert, J.~Andreasson, A.~Rocker, O.~Jönsson, M.~Svenda, S.~Stern,
  K.~Nass, R.~Andritschke, C.~D. Schroeter, F.~Krasniqi, M.~Bott, K.~E.
  Schmidt, X.~Wang, I.~Grotjohann, J.~M. Holton, T.~R.~M. Barends, R.~Neutze,
  S.~Marchesini, R.~Fromme, S.~Schorb, D.~Rupp, M.~Adolph, T.~Gorkhover,
  I.~Andersson, H.~Hirsemann, G.~Potdevin, H.~Graafsma, B.~Nilsson, and
  J.~C.~H. Spence,
\newblock Nature {\bf 470}, 73 (2011).

\bibitem{Seibert}
M.~M. Seibert, T.~Ekeberg, F.~R. N.~C. Maia, M.~Svenda, J.~Andreasson,
  O.~Jonsson, D.~Odic, B.~Iwan, A.~Rocker, D.~Westphal, M.~Hantke, D.~P.
  DePonte, A.~Barty, J.~Schulz, L.~Gumprecht, N.~Coppola, A.~Aquila, M.~N.
  Liang, T.~A. White, A.~Martin, C.~Caleman, S.~Stern, C.~Abergel, V.~Seltzer,
  J.~M. Claverie, C.~Bostedt, J.~D. Bozek, S.~Boutet, A.~A. Miahnahri,
  M.~Messerschmidt, J.~Krzywinski, G.~Williams, K.~O. Hodgson, M.~J. Bogan,
  C.~Y. Hampton, R.~G. Sierra, D.~Starodub, I.~Andersson, S.~Bajt,
  M.~Barthelmess, J.~C.~H. Spence, P.~Fromme, U.~Weierstall, R.~Kirian,
  M.~Hunter, R.~B. Doak, S.~Marchesini, S.~P. Hau-Riege, M.~Frank, R.~L.
  Shoeman, L.~Lomb, S.~W. Epp, R.~Hartmann, D.~Rolles, A.~Rudenko, C.~Schmidt,
  L.~Foucar, N.~Kimmel, P.~Holl, B.~Rudek, B.~Erk, A.~Homke, C.~Reich,
  D.~Pietschner, G.~Weidenspointner, L.~Struder, G.~Hauser, H.~Gorke,
  J.~Ullrich, I.~Schlichting, S.~Herrmann, G.~Schaller, F.~Schopper, H.~Soltau,
  K.~U. Kuhnel, R.~Andritschke, C.~D. Schroter, F.~Krasniqi, M.~Bott,
  S.~Schorb, D.~Rupp, M.~Adolph, T.~Gorkhover, H.~Hirsemann, G.~Potdevin,
  H.~Graafsma, B.~Nilsson, H.~N. Chapman, and J.~Hajdu,
\newblock Nature {\bf 470}, 78 (2011).

\bibitem{chapman2010}
H.~N. Chapman and K.~A. Nugent,
\newblock Nature Photonics {\bf 4}, 833 (2010).

\bibitem{abbey2011}
B.~Abbey, L.~W. Whitehead, H.~M. Quiney, D.~J. Vine, G.~A. Cadenazzi, C.~A.
  Henderson, K.~A. Nugent, E.~Balaur, C.~T. Putkunz, A.~G. Peele, G.~J.
  Williams, and I.~McNulty,
\newblock Nature Photonics {\bf 5}, 420 (2011).

\bibitem{miao1999}
J.~Miao, P.~Charalambous, J.~Kirz, and D.~Sayre,
\newblock Nature {\bf 400}, 342 (1999).

\bibitem{zuo2003}
J.~M. Zuo, I.~Vartanyants, M.~Gao, R.~Zhang, and L.~A. Nagahara,
\newblock Science {\bf 300}, 1419 (2003).

\bibitem{emma2}
P.~Emma, R.~Akre, J.~Arthur, R.~Bionta, C.~Bostedt, J.~Bozek, A.~Brachmann,
  P.~Bucksbaum, R.~Coffee, F.~J. Decker, Y.~Ding, D.~Dowell, S.~Edstrom,
  A.~Fisher, J.~Frisch, S.~Gilevich, J.~Hastings, G.~Hays, P.~Hering, Z.~Huang,
  R.~Iverson, H.~Loos, M.~Messerschmidt, A.~Miahnahri, S.~Moeller, H.~D. Nuhn,
  G.~Pile, D.~Ratner, J.~Rzepiela, D.~Schultz, T.~Smith, P.~Stefan,
  H.~Tompkins, J.~Turner, J.~Welch, W.~White, J.~Wu, G.~Yocky, and J.~Galayda,
\newblock Nature Photonics {\bf 4}, 641 (2010).

\bibitem{ishikawa2012}
T.~Ishikawa, H.~Aoyagi, T.~Asaka, Y.~Asano, N.~Azumi, T.~Bizen, H.~Ego,
  K.~Fukami, T.~Fukui, Y.~Furukawa, S.~Goto, H.~Hanaki, T.~Hara, T.~Hasegawa,
  T.~Hatsui, A.~Higashiya, T.~Hirono, N.~Hosoda, M.~Ishii, T.~Inagaki,
  Y.~Inubushi, T.~Itoga, Y.~Joti, M.~Kago, T.~Kameshima, H.~Kimura,
  Y.~Kirihara, A.~Kiyomichi, T.~Kobayashi, C.~Kondo, T.~Kudo, H.~Maesaka, X.~M.
  Marechal, S.~Masuda, T.and~Matsubara, T.~Matsumoto, T.~Matsushita, S.~Matsui,
  M.~Nagasono, N.~Nariyama, H.~Ohashi, T.~Ohata, T.~Ohshima, S.~Ono, Y.~Otake,
  C.~Saji, T.~Sakurai, T.~Sato, K.~Sawada, T.~Seike, K.~Shirasawa, T.~Sugimoto,
  S.~Suzuki, S.~Takahashi, H.~Takebe, K.~Takeshita, K.~Tamasaku, H.~Tanaka,
  R.~Tanaka, T.~Tanaka, T.~Togashi, K.~Togawa, A.~Tokuhisa, H.~Tomizawa,
  K.~Tono, S.~K. Wu, M.~Yabashi, M.~Yamaga, A.~Yamashita, K.~Yanagida,
  C.~Zhang, T.~Shintake, H.~Kitamura, and N.~Kumagai,
\newblock Nature Photonics {\bf 6}, 540 (2012).

\bibitem{krausz}
F.~Krausz and M.~Ivanov,
\newblock Rev. Mod. Phys. {\bf 81}, 163 (2009).

\bibitem{bucksbaum2007}
P.~H. Bucksbaum,
\newblock Science {\bf 317}, 766 (2007).

\bibitem{corkum2007}
P.~B. Corkum and F.~Krausz,
\newblock Nature Physics {\bf 3}, 381 (2007).

\bibitem{breidbach2003}
J.~Breidbach and L.~S. Cederbaum,
\newblock J. Chem. Phys. {\bf 118}, 3983 (2003).

\bibitem{kuleff2005}
A.~I. Kuleff, J.~Breidbach, and L.~S. Cederbaum,
\newblock J. Chem. Phys. {\bf 123}, 044111 (2005).

\bibitem{remacle}
F.~Remacle and R.~D. Levine,
\newblock Proc. Natl. Acad. Sci. U.S.A {\bf 103}, 6793 (2006).

\bibitem{dutoi2011}
A.~D. Dutoi and L.~S. Cederbaum,
\newblock J. Phys. Chem. Lett. {\bf 2}, 2300 (2011).

\bibitem{benedick2012}
A.~J. Benedick, J.~G. Fujimoto, and F.~X. K{\"a}rtner,
\newblock Nature Photonics {\bf 6}, 97 (2012).

\bibitem{Emma1}
P.~Emma, K.~Bane, M.~Cornacchia, Z.~Huang, H.~Schlarb, G.~Stupakov, and
  D.~Walz,
\newblock Phys. Rev. Lett. {\bf 92}, 74801 (2004).

\bibitem{Zholents}
A.~A. Zholents and W.~M. Fawley,
\newblock Phys. Rev. Lett. {\bf 92}, 224801 (2004).

\bibitem{tanaka2013}
T.~Tanaka,
\newblock Phys. Rev. Lett. {\bf 110}, 084801 (2013).

\bibitem{kumar2013attosecond}
S.~Kumar, H.~S. Kang, and D.~E. Kim,
\newblock Applied Sciences {\bf 3}, 251 (2013).

\bibitem{dixit2012}
G.~Dixit, O.~Vendrell, and R.~Santra,
\newblock Proc. Natl. Acad. Sci. U.S.A. {\bf 109}, 11636 (2012).

\bibitem{dixit2013jcp}
G.~Dixit and R.~Santra,
\newblock J. Chem. Phys. {\bf 138}, 134311 (2013).

\bibitem{dixit2013prl}
G.~Dixit, J.~M. Slowik, and R.~Santra,
\newblock Phys. Rev. Lett. {\bf 110}, 137403 (2013).

\bibitem{abbamonte2004imaging}
P.~Abbamonte, K.~D. Finkelstein, M.~D. Collins, and S.~M. Gruner,
\newblock Phys. Rev. Lett. {\bf 92}, 237401 (2004).

\bibitem{abbamonte2008dynamical}
P.~Abbamonte, T.~Graber, J.~P. Reed, S.~Smadici, C.~L. Yeh, A.~Shukla, J.~P.
  Rueff, and W.~Ku,
\newblock Proc. Natl. Acad. Sci. U.S.A. {\bf 105}, 12159 (2008).

\bibitem{abbamonte2010ultrafast}
P.~Abbamonte, G.~C.~L. Wong, D.~G. Cahill, J.~P. Reed, R.~H. Coridan, N.~W.
  Schmidt, G.~H. Lai, Y.~I. Joe, and D.~Casa,
\newblock Advanced Materials {\bf 22}, 1141 (2010).

\bibitem{abbamonte2009inhomogeneous}
P.~Abbamonte, J.~P. Reed, Y.~I.~Joe, Y.~Gan, and D.~Casa,
\newblock Phys. Rev. B {\bf 80}, 054302 (2009).

\bibitem{abbamonte2012standingwaves}
Y.~Gan, A.~Kogar, and P.~Abbamonte, 
\newblock Chem. Phys. {\bf 414}, 160 (2013).

\bibitem{craig1984}
D.~P. Craig and T.~Thirunamachandran,
\newblock {\em Molecular Quantum Electrodynamics},
\newblock Academic Press, London, 1984.

\bibitem{haverkort2007nonresonant}
M.~W. Haverkort, A.~Tanaka, L.~H. Tjeng, and G.~A. Sawatzky,
\newblock Phys. Rev. Lett. {\bf 99}, 257401 (2007).

\bibitem{loudon1983}
R.~Loudon,
\newblock {\em The Quantum Theory of Light},
\newblock Oxford University Press, Oxford, 1983.

\bibitem{glauber1963}
R.~J. Glauber,
\newblock Phys. Rev. {\bf 130}, 2529 (1963).

\bibitem{henriksen}
N.~E. Henriksen and K.~B. Moller,
\newblock J. Phys. Chem. B {\bf 112}, 558 (2008).

\bibitem{tanaka}
S.~Tanaka, V.~Chernyak, and S.~Mukamel,
\newblock Phys. Rev. A {\bf 63}, 63405 (2001).

\bibitem{schulke2007electron}
W.~Sch{\"u}lke,
\newblock {\em Electron dynamics by inelastic X-ray scattering},
\newblock Oxford University Press Oxford, UK, 2007.

\bibitem{van1954correlations}
L.~Van~Hove,
\newblock Phys. Rev. {\bf 95}, 249 (1954).

\bibitem{petrillo}
C.~Petrillo and F.~Sacchetti,
\newblock Phys. Rev. B {\bf 51}, 4755 (1995).

\bibitem{schuelke1995}
W.~Sch{\"u}lke, J.~R.~Schmitz, H.~Schulte-Schrepping and A.~Kaprolat
\newblock Phys. Rev. B {\bf 52}, 11721 (1995).

\bibitem{watanabe}
N.~Watanabe, H.~Hayashi, Y.~Udagawa, S.~Ten-no and S.~Iwata
\newblock J. Chem. Phys. {\bf 108}, 4545 (1998).

\bibitem{fetter1971}
A.~L. Fetter and J.~D. Walecka,
\newblock {\em Quantum Theory of Many-Particle Systems},
\newblock McGraw-Hill, Boston, 1971.

\end{thebibliography}

\end{document}